\documentclass[twocolumn,prl,showpacs,aps,superscriptaddress]{revtex4-2}

\usepackage{multirow}
\usepackage{graphicx} 
\usepackage{color}
\usepackage[latin1]{inputenc}
\usepackage{enumitem}
\usepackage[french]{babel}
\usepackage{ulem}

\begin{document}

\date{\today}

\title{Cuprates phase diagram deduced from magnetic susceptibility: \\ what is the `true' pseudogap line?}

\author{Yves Noat}

\affiliation{Institut des Nanosciences de Paris, CNRS, UMR 7588 \\
Sorbonne Universit\'{e}, Facult\'{e} des Sciences et Ing\'{e}nierie, 4 place
Jussieu, 75005 Paris, France}

\author{Alain Mauger}

\affiliation{Institut de Min\'{e}ralogie, de Physique des Mat\'{e}riaux et
de Cosmochimie, CNRS, UMR 7590,Sorbonne Universit\'{e}, Facult\'{e} des Sciences et Ing\'{e}nierie, 4 place
Jussieu, 75005 Paris, France}

\author{Minoru Nohara}

\affiliation{Department of Quantum Matter, Hiroshima University,
1--3--1 Kagamiyama, Higashi-Hiroshima, Japan 739-8530}

\author{Hiroshi Eisaki}

\affiliation{Research Institute for Advanced Electronics and
Photonics (RIAEP), National Institute of Advanced Industrial Science
and Technology (AIST), Tsukuba, Ibaraki 305-8568, Japan}

\author{Shigeyuki Ishida}

\affiliation{Research Institute for Advanced Electronics and
Photonics (RIAEP), National Institute of Advanced Industrial Science
and Technology (AIST), Tsukuba, Ibaraki 305-8568, Japan}

\author{William Sacks}

\affiliation{Institut de Min\'{e}ralogie, de Physique des Mat\'{e}riaux et
de Cosmochimie, CNRS, UMR 7590,Sorbonne Universit\'{e}, Facult\'{e} des Sciences et Ing\'{e}nierie, 4 place
Jussieu, 75005 Paris, France}


\begin{abstract}
Two contradictory phase diagrams have dominated the literature of
high-$T_c$ cuprate superconductors. Does the pseudogap line cross
the superconducting $T_c$-dome or not? To answer, we have revisited
the experimental magnetic susceptibility and knight shift of four
different compounds, La$_{1-x}$Sr$_x$CuO$_4$,
Bi$_2$Sr$_2$Ca$_{1-x}$Y$_x$Cu$_2$O$_8$,
Bi$_2$Sr$_2$CaCu$_2$O$_{8+y}$, YBa$_2$Cu$_3$O$_{6+y}$, as a function
of temperature and doping. The susceptibility can be described by
the same function for all materials, having a magnetic and an
electronic contributions. The former is the 2D antiferromagnetic
(AF) square lattice response, with a characteristic temperature of
magnetic correlations $T_{max}$. The latter is the `Pauli' term,
revealing the gap opening in the electronic density of states at the
pseudogap temperature $T^*$.

From precise fits of the data, we find that $T_{max}(p)$ decreases
linearly as a function of doping ($p$) over a wide range, but
saturates abruptly in the overdoped regime. Concomitantly, $T^*(p)$
is {\it linear and tangent} to the dome, either crossing or
approaching $T_{max}(p)$ at the top of the dome, indicating a
qualitative change of behavior from underdoped to overdoped regimes.

Contrary to the idea that the pseudogap terminates just above
optimal doping, our analysis suggests that the gap exists throughout
the phase diagram. It is consistent with a pseudogap due to hole
pairs, or `pairons', above $T_c$.  We conclude that $T_{max}$,
reflecting the AF magnetic correlations, has often been
misinterpreted as the pseudogap temperature $T^*$.
\end{abstract}

\maketitle

\subsection{Introduction}

\begin{figure}
\includegraphics[width=8.4 cm]{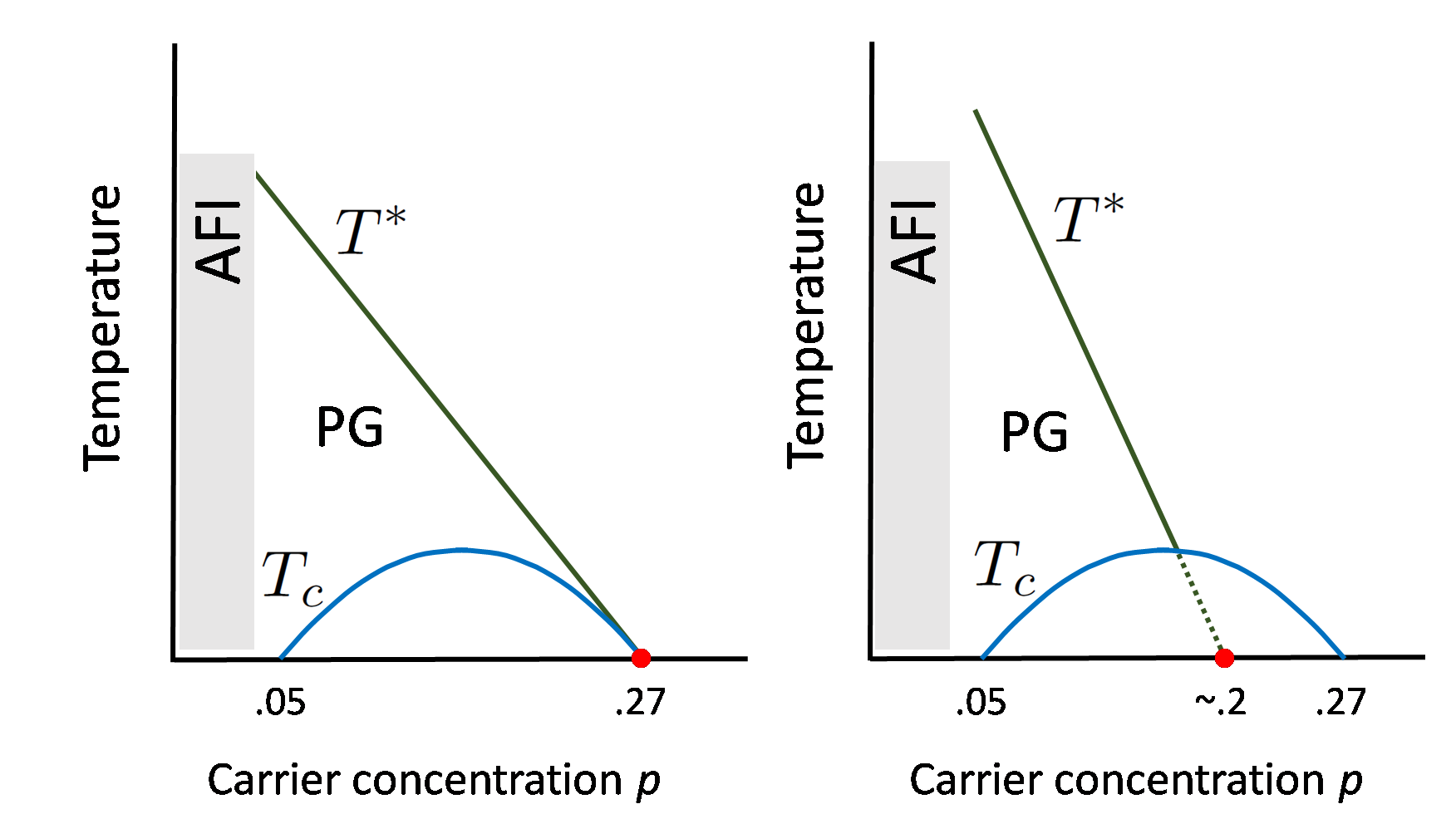}
\caption{(Color online) Schematic of the two generic phase diagrams: Left panel: A pseudogap exists for all doping and ending at the maximum doping $p_c=$0.27. This phase diagram is generally deduced from direct probes such as tunneling and photoemission spectroscopy. Right panel: The pseudogap line crosses the superconducting dome, ending at the critical doping $p_c\approx$0.2. This class of phase diagrams is generally deduced from indirect probes, such as transport, specific heat or magnetic susceptibility measurements.} \label{Fig_Phase diagrams}
\end{figure}
Since the discovery of cuprates by Bednorz, and M\"{u}ller \cite{ZPhys_Bednorz1986} in 1986, the phase diagram of high-$T_c$ superconductors remains a puzzle. In the underdoped regime a gap persists at the Fermi level above the critical temperature, called the pseudogap (PG) (see Ref.\,\cite{Rep_ProgPhys_Timusk1999} for a review). First discovered by NMR \cite{PRL_Warren1989,PRL_Alloul1989}, it was rapidly confirmed by optical conductivity \cite{PRL_Rotter1991,PRL_Homes1993}, neutron scattering \cite{PRL_Shirane1989,PhysicaB_Rossat1991}, transport \cite{PRL_Ito1993,PRL_Bucher1993,PhysicaC_Batlogg1994,PhysicaC_Watanabe1997}, specific heat \cite{PhysicaC_Loram1994}, tunneling \cite{PhysicaC_Tao1997,PRL_renner1998_T} and photoemission spectroscopies \cite{Nat_Ding1996,PhysicaC_Loeser1996}.

The pseudogap is one of the key ingredients distinguishing high-$T_c$ cuprates from conventional superconductors, which are successfully described by the Bardeen-Cooper-Schrieffer (BCS) theory \cite{PR_BCS1957}. Indeed, BCS superconductors are characterized by a gap in the quasiparticle excitation spectrum that closes at the critical temperature concomitantly with the disappearance of superconducting coherence.

While the existence of a pseudogap phase above $T_c$ is well
established, its relationship to the superconducting state is still
strongly debated, as discussed in detail by Kordyuk in Ref.
\cite{LowTempPhys_Kordyuk2015}. Two main avenues have emerged in the
literature to understand this issue:

i) The pseudogap is a precursor of superconductivity, with incoherent preformed pairs existing above $T_c$.

ii) The pseudogap is linked to a competing order such as a spin density wave or charge order.

According to the hypothesis (i), the pseudogap exists for any doping value in the range where $T_c$ does not vanish, while according to (ii), the pseudogap exists only below some lower critical value of the hole concentration. In order to address this unresolved issue, it is crucial to know the temperature at which the gap in the electronic density of states (DOS) opens, as a function of carrier concentration.

We define $T^*$, and maintain this definition throughout, as the temperature at which a quasiparticle gap at the antinodal point  $(0,\pi)$ in angle-resolved photoemission spectroscopy (ARPES) or a gap at the Fermi level in tunneling, vanishes with rising temperature. The PG temperature $T^*$ can be measured {\it directly} by electronic spectroscopic probes, such as tunneling or photoemission spectroscopies (see Ref. \cite{Revmod_Fisher2007} and \cite{Revmod_Damascelli2003} for comprehensive reviews), but only {\it indirectly} through the resistivity, specific heat or magnetic susceptibility. In the latter cases, the determination of $T^*$ can be complex, as we shall discuss in this paper for the magnetic susceptibility, since the gap in the electronic DOS must be inferred using an appropriate theory.

Based on the wide variety of measurements mentioned above, two general classes of phase diagrams are readily encountered in the literature (see figure \ref{Fig_Phase diagrams}), each of them in favor of one of the two hypotheses mentioned above.
In the first class (Fig.\,\ref{Fig_Phase diagrams}, left panel), the pseudogap exists for any doping along the superconducting dome \cite{RepProgPhys_Hufner2008}, where the $T^*(p)$ line arrives tangential to the dome. In an alternative phase diagram (Fig.\,\ref{Fig_Phase diagrams}, right panel), the pseudogap is suggested to terminate near the top of the dome or inside the dome \cite{PhysicaC_Konstantinovi2001,PhysicaC_Naqib2003,NatCom_Sterpetti2017}, at a critical value associated with a quantum critical point (see \cite{AnRevCondMat_Proust2019} and Ref. therein).

Not only is there a quantitative difference between the two phase diagrams but also a qualitative one: Indeed, using ARPES, several teams have shown the existence of the pseudogap in the DOS above $T_c$ in the overdoped regime, at least up to $p=0.2$ \cite{PRB_Hashimoto2007,PNAS_Vishik2012} whereas Loram et al. and Naqib et al. report a pseudogap deduced from the magnetic susceptibility \cite{PhysicaC_Naqib2007,ScSciTech_Naqib2008}, the resistivity \cite{PhysicaC_Naqib2003} and the specific heat \cite{JPhysChemSol_Loram1998,JphysChem_Loram2001}, vanishing at $p=$0.19.

In this article, we revisit the magnetic susceptibility of high--$T_c$ cuprates and show that a proper analysis allows to reconcile the contradictory phase diagrams of the literature. Our results show that there is not one, but {\it two} characteristic temperatures present in the phase diagram. The first one is $T^*$, unambiguously defined as the temperature at which a gap in the electronic DOS opens at the Fermi level in the antinodal direction. The second is $T_{max}$, the characteristic temperature of 2D antiferromagnetic correlations. It is defined experimentally as the temperature of the maximum in the magnetic susceptibility.

Our analysis points to a the pseudogap in the DOS, following $T^*$, existing for all doping values along the superconducting dome. The temperature dependence of the DOS is due to the excitation of hole pairs or pairons \cite{EPL_Sacks2017,EPL_Noat2019} above $T_c$, and their dissociation into quasiparticles \cite{EPJB_Sacks2016,Jphys_Sacks2018}. The `so-called' pseudogap deduced in previous works from the magnetic susceptibility \cite{PhysicaC_Naqib2007,ScSciTech_Naqib2008,PhysicaB_Lopes2018} is in fact  $T_{max}$, the characteristic temperature of magnetic correlations. It is clearly distinct from $T^*$, that is the onset temperature of a gap in the electronic DOS at the Fermi level. While both temperatures depend on a unique energy scale, the exchange energy $J$, we show in this paper that they are not simply proportional, as suggested in Ref. \cite{JPhysSocJap_Nakano1998}.

\subsection{Magnetic susceptibility of high-$T_c$ cuprates}

\begin{figure}
\includegraphics[width=9.2 cm]{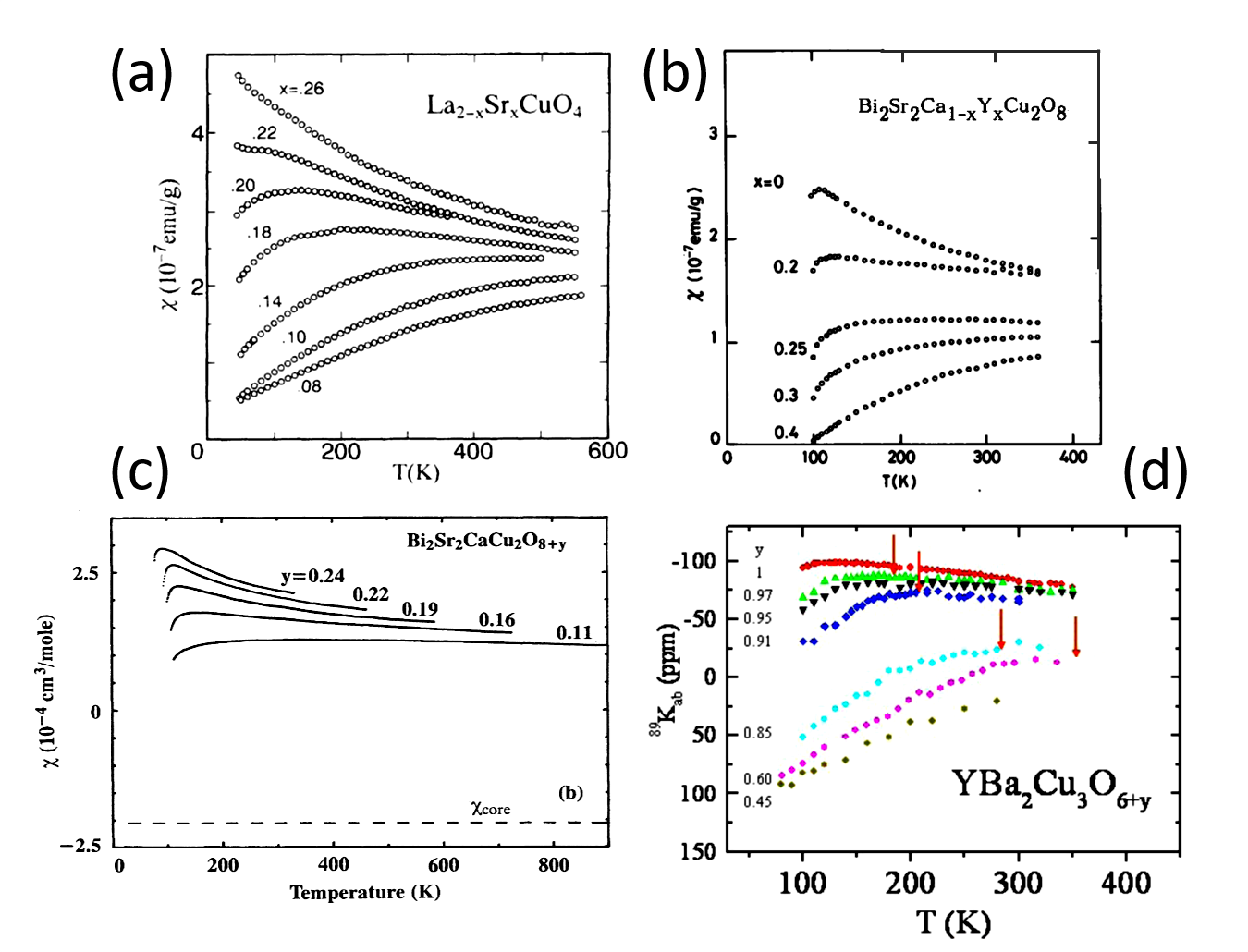}
\caption{(Color online) Experimental susceptibility as function of temperature and doping measured for different materials in the cuprate family: a) La$_{2-x}$Sr$_x$CuO$_4$ by Nakano et al.\cite{PRB_Nakano1994}, b) Bi$_2$Sr$_2$Ca$_{1-x}$Y$_x$Cu$_2$O$_8$ by Oda et al. \cite{SolstatCom_Oda1990}, c) Bi$_2$Sr$_2$CaCu$_2$O$_{8+y}$ by Allgeier et al. \cite{PRL_Allgeier1993} d) Knight shift measured in YBa$_2$Cu$_3$O$_{6+y}$ by Alloul et al. \cite{Book_Alloul2016}.} \label{Fig_Data}
\end{figure}

The magnetic susceptibility $\chi(T)$ of cuprates has been extensively studied as function of temperature and doping for different materials \cite{PRL_Johnston1989,PRB_Torrance1989,PRB_Takagi1989, PhysicaC_Yoshizaki1990,PRB_Oda1990,PhysicaC_Oda1997,PRB_Wakimoto2005}. We have chosen representative materials and measurements with a wide doping range. We have focused attention on La$_{1-x}$Sr$_x$CuO$_4$ (LSCO) by Nakano et al. \cite{PRB_Nakano1994}, Y-doped  Bi$_2$Sr$_2$Ca$_{1-x}$Y$_x$Cu$_2$O$_8$ (Y-BSCCO) by Oda et al. \cite{SolstatCom_Oda1990}, oxygen doped Bi$_2$Sr$_2$CaCu$_2$O$_{8+y}$ (BSCCO) by Allgeier et al. \cite{PRL_Allgeier1993}. We have also analyzed the Knight shift measured in YBa$_2$Cu$_3$O$_{6+y}$ (YBCO) by Alloul et al. \cite{Book_Alloul2016}, (see Fig.\ref{Fig_Data}).

Clearly, even before a detailed analysis, one observes similar trends in the data: at low hole doping, i.e. in the underdoped regime,  $\chi(T)$ is a smooth increasing function of temperature. However, at intermediate doping, i.e. close to the optimal doping value $p=0.16$, $\chi(T)$ develops a pronounced maximum at a characteristic temperature, which is followed by a power law decay. This maximum monotically decreases towards $T_c$ in the overdoped regime.


In the underdoped case, where the susceptibility increases with temperature, it is tempting to offer an immediate explanation based on the electronic DOS.
Since in a metal the electronic Pauli susceptibility is independent of temperature, one tempting interpretation is to attribute this behavior to a gap in the electronic DOS above $T_c$. In this approach, a pseudogap would exist in the doping range where the susceptibility decreases upon cooling, i.e. roughly below optimal doping.

This is the spirit of the work of Naqib et al., who report in several articles the measurement of the magnetic susceptibility as  function of temperature and doping in LSCO and YBCO \cite{PhysicaC_Naqib2007,ScSciTech_Naqib2008}. To analyse their data, they assumed that $\chi(T)$ is essentially due to the temperature dependence of the electronic DOS at the Fermi level. The reduction of $\chi(T)$ observed at low temperature in the underdoped regime was then explained using a temperature-independent gap at the Fermi level. The temperature dependence of $\chi(T)$  was then attributed to the thermal electron--hole excitations through this gap, like in any semiconductor.

From the fit of their data, they deduced an energy gap $E_g$ that varies linearly with $p$, vanishing at a critical value $p=$0.19. Their approach provides a satisfactory explanation for YBCO, given the good quality of the fits. Their temperature scale $E_g/k_B$ gives a phase diagram belonging to the class of Fig.\,\ref{Fig_Phase diagrams}, right panel.

However, the approach no longer works for LSCO, since for intermediate hole doping, the
 susceptibility $\chi(T)$ develops a maximum, followed by a power law decrease as a function of temperature. Moreover, as in a more recent paper \cite{SciRep_Tallon2020}, the gap energy $E_g$ shows no sign of closing up to high temperatures ($T\sim$400$K$), but such a rigid gap is difficult to justify in a metallic system.
Moreover, above optimum doping, the concave nature of $\chi(T)$ becomes even more pronounced and its shape as a function temperature can no longer be described by a gap in the electronic DOS.

\subsection{Contribution of the AF 2D lattice to the susceptibility}

In this paper we take a different approach that overcomes these discrepensies and gives an accurate description of the magnetic susceptibility. It includes the magnetic contribution of the 2D lattice which has been very well established in the literature of cuprates in the 1990's. Indeed, the overall shape of the magnetic susceptibility has been convincingly attributed to the response of the AF CuO planes, not included by Naqib et al. \cite{PhysicaC_Naqib2007,ScSciTech_Naqib2008}.

In a pioneering work, Johnston \cite{PRL_Johnston1989} has shown that the magnetic susceptibility in LSCO is dominated by the magnetic contribution of the CuO square lattice. This approach was extensively revisited by Nakano et al.\,\cite{PRB_Nakano1994} and independently by other authors \cite{PRL_Allgeier1993,PRB_Wakimoto2005}. 
In particular, the magnetic response of a 2D Heisenberg antiferromagnetic square lattice has been calculated in the literature. Given the Heisenberg Hamiltonian $H=\sum_{i,j}^{}JS_i.S_j$, where the sum runs over all pairs of nearest neighbors i and j, the general form for the susceptibility of the 2D AF square lattice is given by Lines \cite{JphysChemSol_Lines1970}:
\begin{equation}
\frac{Ng^2\mu_b^2}{\chi(T) J}=3\vartheta+\sum_{n=1}^{\infty}\frac{C_n}{\vartheta^{n-1}}
\label{Equa_Chi_AF_2D}
\end{equation}
where $\vartheta=k_BT/\left[JS(S+1)\right]$ with $k_B$ being the Boltzmann constant, $N$ the number of spins, $g$ the Land\'e $g$-factor, $\mu_B$ the Bohr magneton. The coefficients $C_n$ in the series of Eq. \ref{Equa_Chi_AF_2D} are known and tabulated in Ref.\,\cite{JphysChemSol_Lines1970} for different spin value.

\begin{figure}
\hspace*{-5 mm}\includegraphics[width=10.cm]{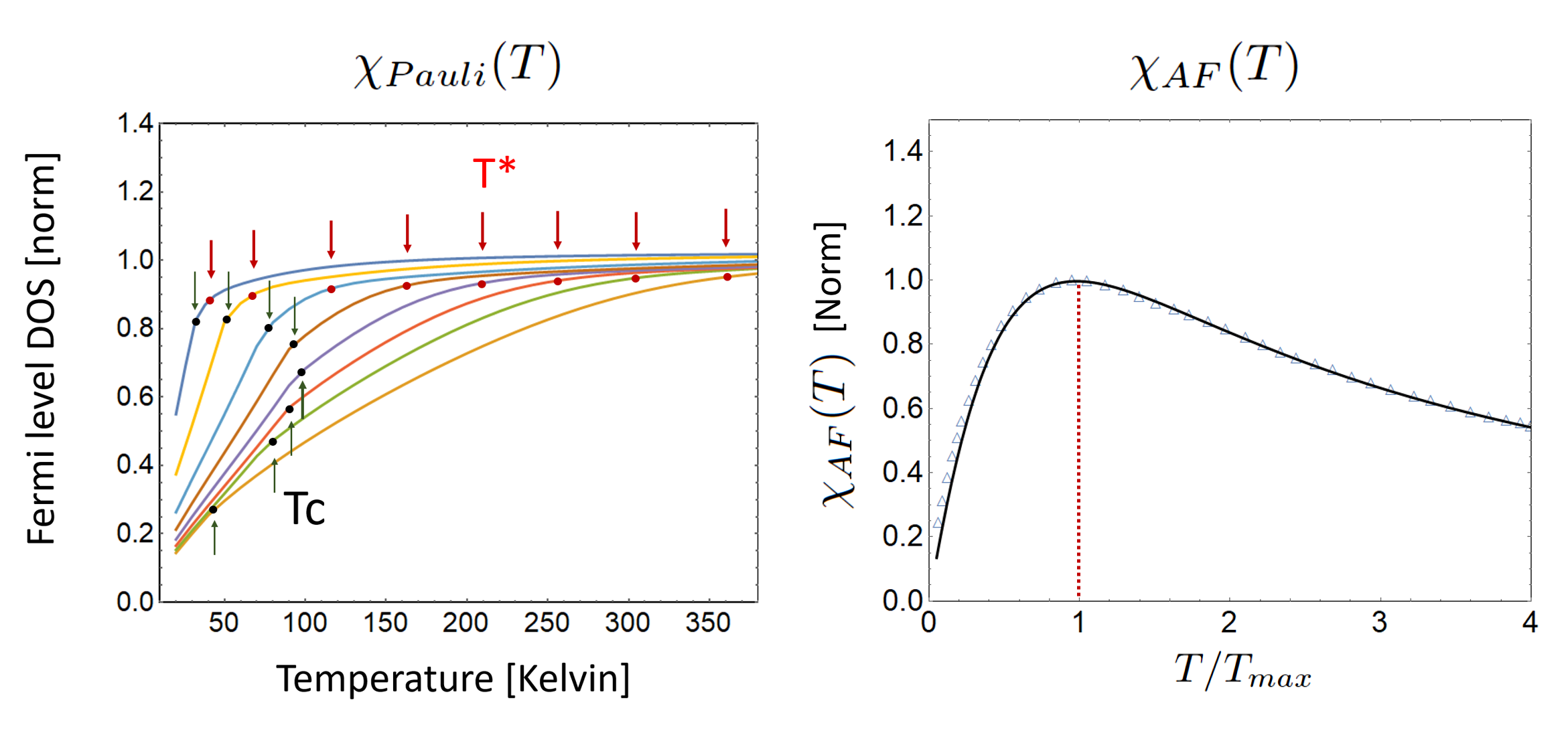}
\caption{(Color online) a) Temperature dependent density of states
at the Fermi level for different hole doping, calculated in the
pairon model. It takes into account the convolution with the
Fermi-Dirac distribution, as in formula \ref{Equa_Chi_Pauli}. b)
Magnetic contribution of the susceptibility using equation
\ref{Equa_Chimag}, solid line. Open triangles : results of Nakano et
al. in La$_{1-x}$Sr$_x$CuO$_4$ \cite{PRB_Nakano1994}, showing a
perfect match.} \label{Fig_Chimag_DOS}
\end{figure}

Although Eq. \ref{Equa_Chi_AF_2D} was rigorously established for the undoped 2D square lattice, we found that the magnetic part of the doped system is satisfactorily described by the first two terms of the sum in Eq. \ref{Equa_Chi_AF_2D}, leading to:
\begin{equation}
\chi_{AF}(T)=A_{mag}(T+\frac{T_{max}^2}{T}+C)^{-1}
\label{Equa_Chimag}
\end{equation}
where $A_{mag}$, $T_{max}$ and $C$ are doping-dependent parameters. The magnetic part of the susceptibility gives a universal curve (see Fig.\,\ref{Fig_Chimag_DOS}, right panel), having a peak at $T=T_{max}$, the characteristic scale of AF correlations, and a $\chi_{AF}(T)\sim A_{mag}/(T+C)$ Curie or Curie-Weiss behavior at high temperature (i.e. for $T\gg T_{max}$).
Note that $T_{max}$ plays a central role since it reflects the characteristic temperature below which magnetic correlations are important.

In a first approach, one reproduces the measured magnetic susceptibility in LSCO (Fig.\,\ref{Fig_Data}a) by assuming that $\chi(T)$ is given by Eq.\,\ref{Equa_Chimag} with an additional constant arising from the electronic Pauli susceptibility:
\begin{equation}
\chi(T)=\chi_{Pauli}+\chi_{AF}(T)
\end{equation}
From the fits of the data, we extract the doping dependence of $T_{max}$ which  accurately reproduces the results of Nakano et al \cite{PRB_Nakano1994}: $T_{max}(p)$ follows a straight line for a wide doping range, in agreement with early calculations \cite{JAppPhys_Glenister1993}, which extrapolates to zero at a value $p\approx$0.23 at $T=0$. However,  $T_{max}(p)$ does not vanish but saturates in the overdoped regime, suggesting that the magnetism is persistent there.

The case of LSCO, where both $T_c$ and $T^*$ are a factor of two
smaller than in BSCCO and YBCO, allows to explain very
satisfactorily the series of observations in terms of the AF
magnetic contribution. We now focus our attention on the other
cuprates, where a more complete {analysis} is needed.
\begin{figure}
\includegraphics[width=8.4 cm]{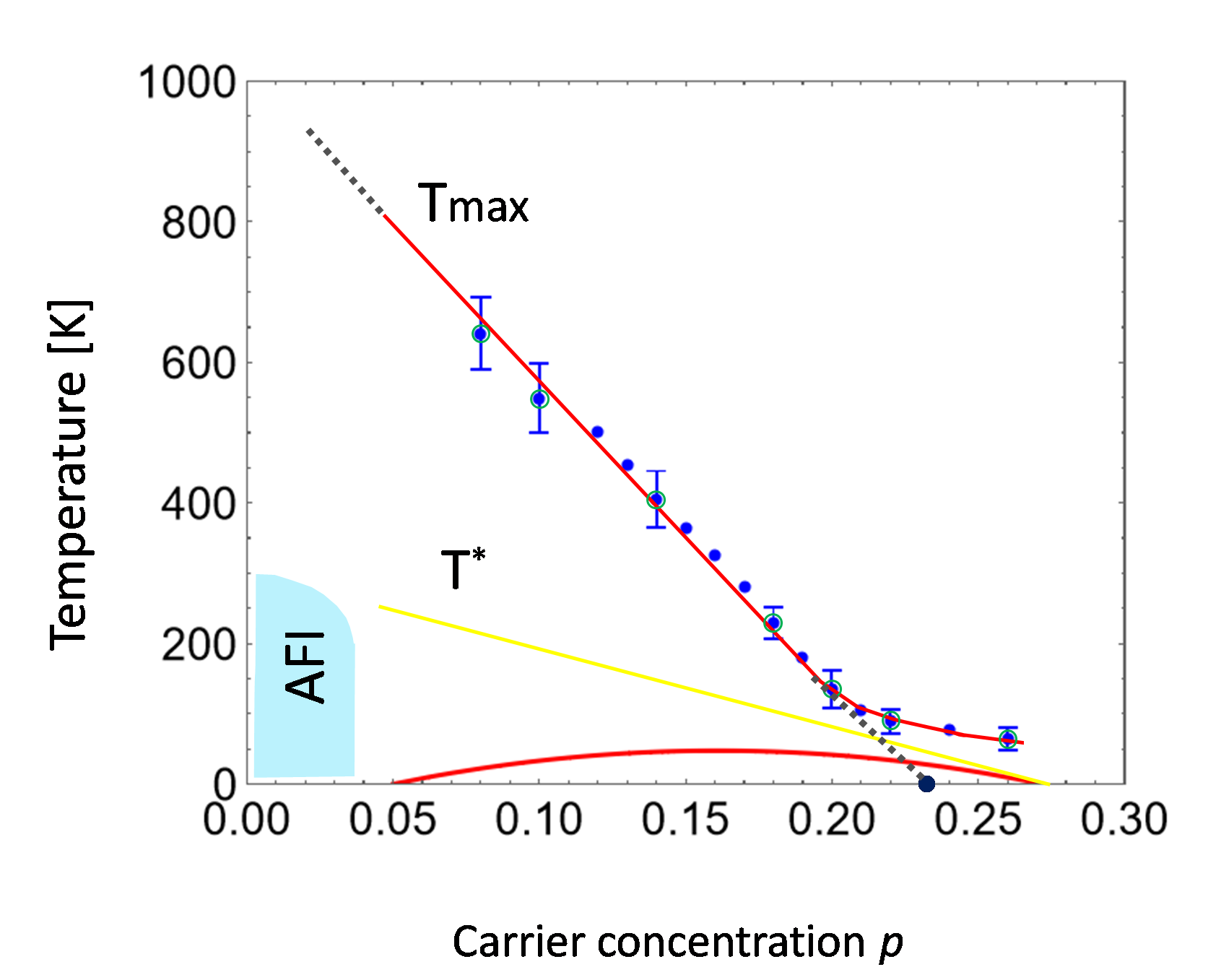}
\caption{(Color online) The magnetic characteristic temperature $T_{max}$ determined from the spin susceptibility of La$_{2-x}$Sr$_x$CuO$_4$ measured by Nakano et al.\cite{PRB_Nakano1994}. Points for $T_{max}$ with error bars are deduced from the fits of Fig.\,\ref{Fig_Data}a; the remaining points are inferred from an interpolation procedure of the data. One can see that the extrapolation of the straight line at $T=0$ is 0.23.} \label{Fig_Tmax LSCO}
\end{figure}

\subsection{General analysis of the susceptibility}

In order to describe the susceptibility of cuprates in a more
general way, we need to extend the model described above. In
particular, one has to include the effect of the pseudogap in the
electronic DOS.

For this purpose, we write the susceptibility as a sum of the following contributions:
\begin{equation}
\chi(T)=\chi_0+\chi_{AF}(T)+\chi_{Pauli}(T)+\chi_{dia}(T)
\label{Equa_Chi_terms}
\end{equation}
The first term is a constant, independent of temperature and doping,
which groups together the atomic core and Van Vleck contributions to
the susceptibility. The second term, $\chi_{AF}(T)$, is the response
of the AF square lattice, as mentioned previously. The third term,
$\chi_{Pauli}(T)$, is the electronic term arising from the
delocalized electrons at the Fermi level. The last term,
$\chi_{dia}(T)$, is the diamagnetic contribution arising from
superconducting currents, relevant close to $T_c$.

\begin{figure}
\includegraphics[width=4.75 cm]{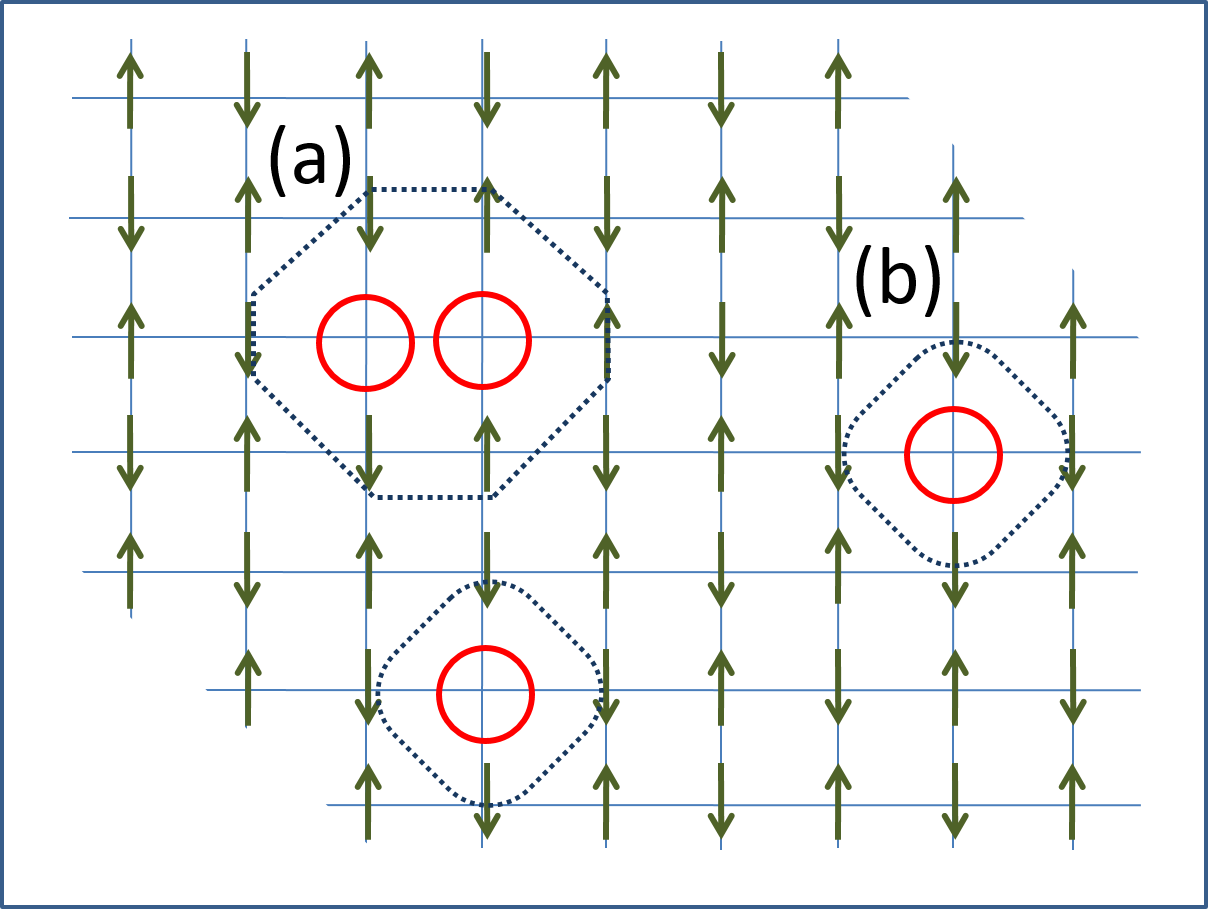}
\caption{(Color online) Schematic diagram of a pairon (a) versus a
hole (b) in their local antiferromagnetic environment, on the scale
of the antiferromagnetic correlation length, $\xi_{AF}$. }
\label{Fig_Pairons}
\end{figure}

We now focus on the electronic term that we will evaluate in the
framework of the pairon model \cite{SciTech_Sacks2015}. In a
previous work, we have proposed that in cuprate superconductors
pairing occurs in an unconventional way, very distinct from the BCS
scenario where Cooper pairs are bound via phonon exchange. In
high-$T_c$ cuprates, hole pairs (which we call `pairons') form
directly as a result of their local antiferromagnetic environment
\cite{EPL_Sacks2017,EPL_Noat2019}, without phonon or magnon exchange
(see Fig.\ref{Fig_Pairons}). This binding mechanism provides an
energy gain of the order of $J$, the AF exchange energy, as
confirmed by early numerical calculations
\cite{PRB_Kaxiras1988,PRB_Bonca1989,PRB_Riera1989,PRB_Hasegawa1989,PRB_Poilblanc1994}.
The characteristic temperature of pairon formation is by definition
$T^*$, directly proportional to $J$
\cite{EPL_Sacks2017,Jphys_Sacks2018}.

Pairons are composite bosons which condense in the collective
superconducting state below $T_c$, as a result of pairon-pairon
interactions, which sets the global phase coherence. In this
approach, two energy scales are thus relevant, the gap $\Delta_p$
(pairon binding energy in the SC state) and $\beta_c$, the
{condensation energy per pair} \cite{SciTech_Sacks2015}. The first
is associated with the pseudogap temperature $T^*$, while the second
is proportional to $T_c$ ($\beta_c\simeq 2.2k_BT_c$). Thus, contrary
to the BCS theory, here the gap is not the order parameter since it
does not vanish at $T_c$.

\begin{figure}
\includegraphics[width=8.0 cm]{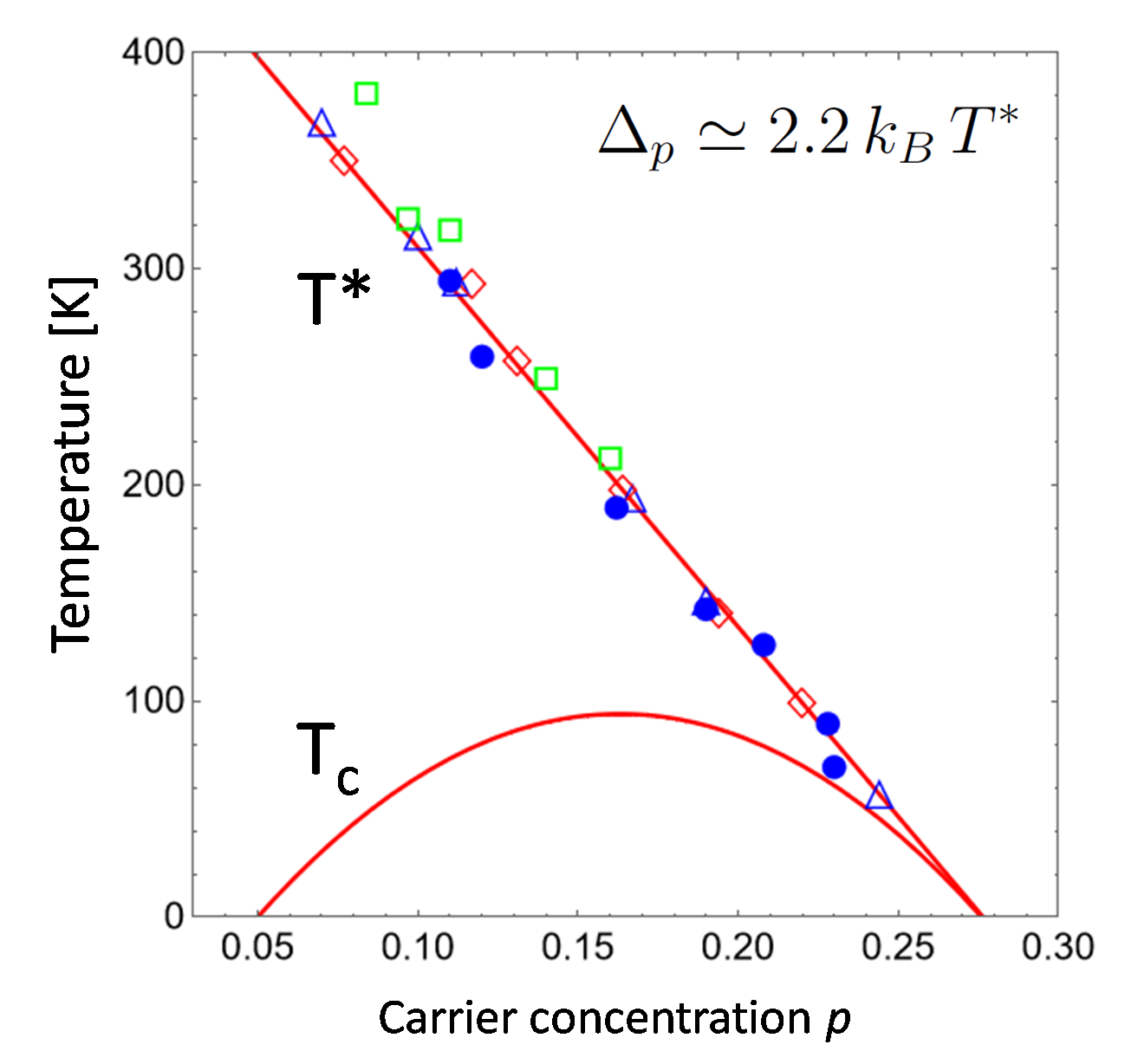}
\caption{(Color online) open triangles and squares: antinodal energy
gap measured at $T=0$ expressed in Kelvin, $\Delta_p/2.2k_B$. Filled
circles: direct temperature measurement of the gap closing at the
higher value $T^*$. The data points are ARPES measurements in BSCCO
taken from the article of Vishik et al. \cite{PNAS_Vishik2012} and
Hashimoto et al. \cite{Nat_Hashimoto2014}.} \label{Fig_AN_gap}
\end{figure}

Both ARPES
\cite{PNAS_Vishik2012,PRB_Hashimoto2007,Nat_Hashimoto2014} and
tunneling
\cite{PRL_renner1998_T,JPhysChemSol_Gomes2008,JphysSocJap_Sekine2016}
measurements have confirmed that the antinodal gap is still present
at the critical temperature and closes at a higher temperature
$T^*$. Furthermore, Fig.\,\ref{Fig_AN_gap} illustrates two
astonishing aspects: first the measured temperature of the closing
of the gap, $T^*$, is directly proportional to the {\it zero
temperature gap}, with a proportionality factor given by the
relation $\Delta_p\approx 2.2k_BT^*$. The factor 2.2 has been
previously determined from detailed fits of ARPES and tunneling
spectra \cite{EPL_Sacks2017,Jphys_Sacks2018}. Secondly, the
dependence of the gap $\Delta_p$ is practically linear with carrier
concentration.

Given that the Pauli contribution to the susceptibility is proportional to the DOS at the Fermi energy, in what follows we calculate this quantity. In a BCS superconductor, above the critical temperature the Pauli susceptibility should be roughly independent of temperature. On the contrary, in the pairon model for cuprates, involving pair excitations and pair dissociations, the DOS at the Fermi energy depends explicitly on temperature up to $\sim T^*$.

More precisely, at finite temperature, pairons are excited out of the condensate following Bose-Einstein statistics and dissociate into quasiparticles, preferentially close to the node, leading to Fermi arcs \cite{Nat_Norman_1998}. This effect, as studied in Ref.\,\cite{Jphys_Sacks2018} gives rise to a temperature-dependent DOS:
\begin{equation}
N_{ex}(E,T)=\sum_in_i(\varepsilon_i,T)\int \frac{d\theta}{2\pi}N^i(E,\Delta_i(\theta))
\label{Equa_Nex_1}
\end{equation}
where $N^i(E,\Delta_i(\theta))$ is the standard angular-dependent quasiparticle DOS, $\varepsilon_i$ are the excited pairon energies, $n_i(\varepsilon_i,T)$ is the number of excited pairons with associated quasiparticles $E_k^i=\sqrt{\epsilon_k^2+\Delta_i^2}$.

Given the density of pair states $P_0(\varepsilon_i)$, one has:
\begin{equation}
n_i(\varepsilon_i,T)=AP_0(\varepsilon_i)f_{BE}(\varepsilon_i)
\end{equation}
where $f_{BE}(\varepsilon)=1/\left(\exp\left(\frac{\varepsilon-\mu_b}{k_BT}\right)-1\right)$ is the Bose-Einstein distribution and where $A$ is a normalization factor.
We further impose particle conservation $\sum_i n_i(\varepsilon_i,T)=n_0$, where $n_0$ is the number of pairs, which determines both the constant $A$ and the chemical potential $\mu(T)$ given that $\mu_b(T)=0$ for $T\leq T_c$.

As in our previous work, the density of pairon excited states is assumed to have a Lorentzian form:
\begin{equation}
P_0(\varepsilon_i)=\frac{\sigma_0^2}{\left[(\varepsilon_i-\beta_c)^2+\sigma_0^2\right]}
\end{equation}
where $\sigma_0$ is the width of the distribution. The evaluation of Eq. \ref{Equa_Nex_1} needs the relation between the boson excitations ($\varepsilon_i$) and the associated quasiparticles ($E_k^i$). We use the relation: $\varepsilon_i=\Delta_i-\Delta_p(T,\theta)$, where $\Delta_p(T,\theta)$ is the standard $d$-wave gap with a smooth and decreasing temperature dependence as in Ref.\,\cite{SolStatCom_Noat2021}. With these considerations, we can write:
\begin{equation}
N_{ex}(E,T)=N_n\sum_in_i(\varepsilon_i,T)\int \frac{d\theta}{2\pi}\frac{E}{\sqrt{E^2-(\varepsilon_i+\Delta_p(T,\theta))^2}}
\label{Equa_Nex}
\end{equation}
where $N_n$ is the normal DOS.

Equation \ref{Equa_Nex} leads to a T-dependent DOS at the Fermi energy $N_{ex}(0,T)$ which varies significantly up to the pseudogap temperature $T^*$.
The associated Pauli susceptibility is then determined by the standard formula:
\begin{equation}
\chi_{Pauli}(T)\propto \int-\frac{\partial f(E,T)}{\partial E}N_{ex}(E,T)dE
\label{Equa_Chi_Pauli}
\end{equation}
where $f(E,T)$ is the Fermi-Dirac distribution.
The precise numerical calculation is plotted for different doping values in Fig. \ref{Fig_Chimag_DOS} (see supplementary materials for further details on the calculation of the DOS).

Finally, close to the SC transition (for $T\lesssim T_c+\Delta T$), the aforementioned diamagnetic current term must be included. We use the very simple form:
\begin{equation}
\chi_{dia}(T)=A_{dia}\exp(-\frac{T-T_c}{\Delta T})
\end{equation}
where $A_{dia}$ is a negative constant and $\Delta T$ characterizes the existence of diamagnetic currents above $T_c$.

\subsection{Fits of the data}

\begin{figure}
\includegraphics[width=9.5 cm]{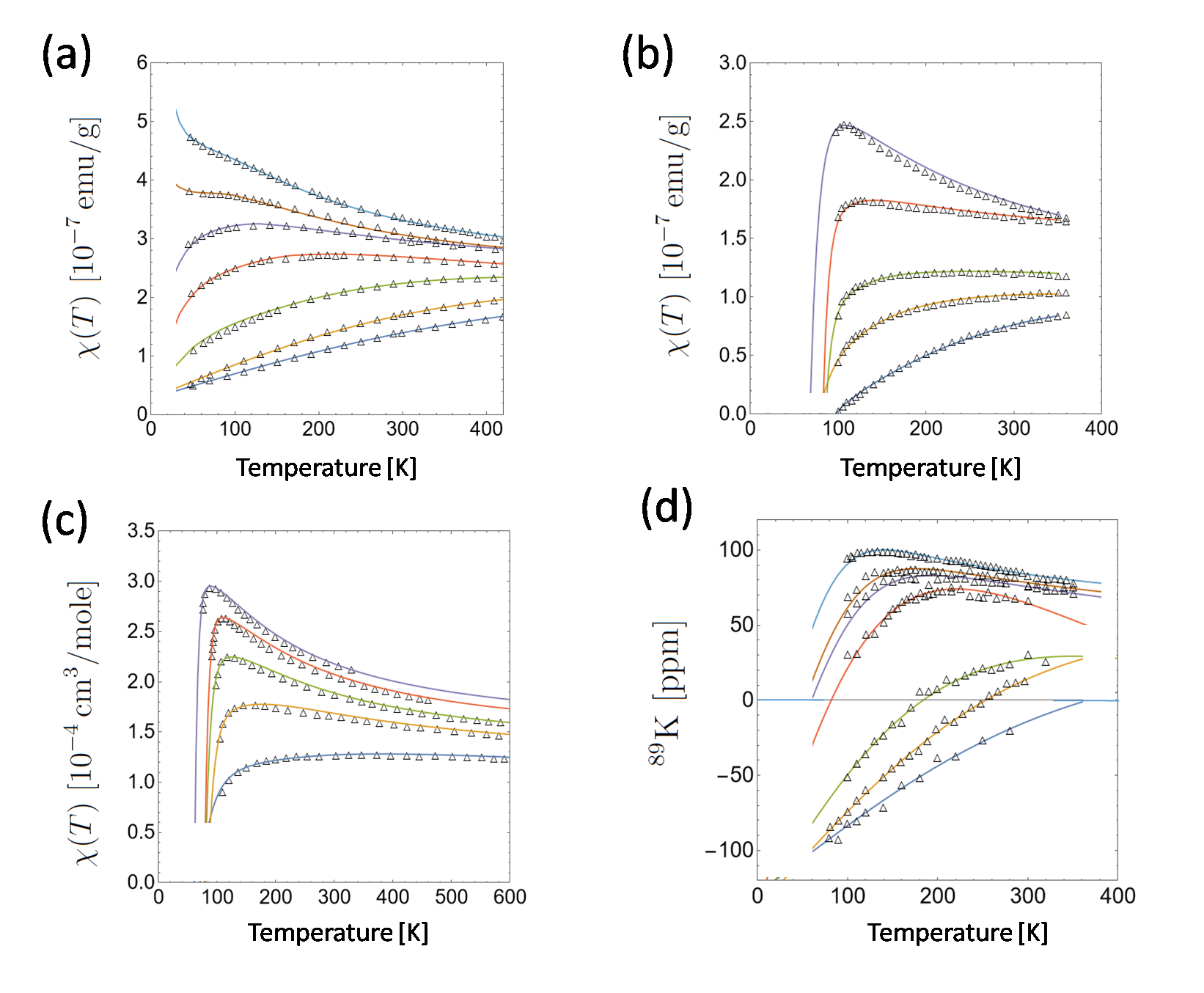}
\caption{(Color online) Colored full lines are the fits, using the analysis described in the text, to the experimental data of Fig.\,\ref{Fig_Data} (dots),  reported for  a) La$_{1-x}$Sr$_x$CuO$_4$ by Nakano et al.\cite{PRB_Nakano1994}, b) Bi$_2$Sr$_2$Ca$_{1-x}$Y$_x$Cu$_2$O$_8$ by Oda et al. \cite{SolstatCom_Oda1990}, c) Bi$_2$Sr$_2$CaCu$_2$O$_{8+y}$ by Allgeier et al. \cite{PRL_Allgeier1993}. d) Knight shift measured in YBa$_2$Cu$_3$O$_{6+y}$ \cite{Book_Alloul2016}.} \label{Fig_Fits}
\end{figure}
We have fitted the experimental susceptibility for the four materials (LSCO, Y-BSCCO, BSCCO, YBCO). Results are shown in Fig. \ref{Fig_Fits}. Let us point out that the same function was used with success for the four different compounds, apart from a small supplementary `Curie' paramagnetic term ($\propto 1/T$) term  in the fit for LSCO in the overdoped regime \cite{CzechJphys_Nakano1996}.
\begin{figure*}
\centering
\includegraphics[height=5.0 cm]{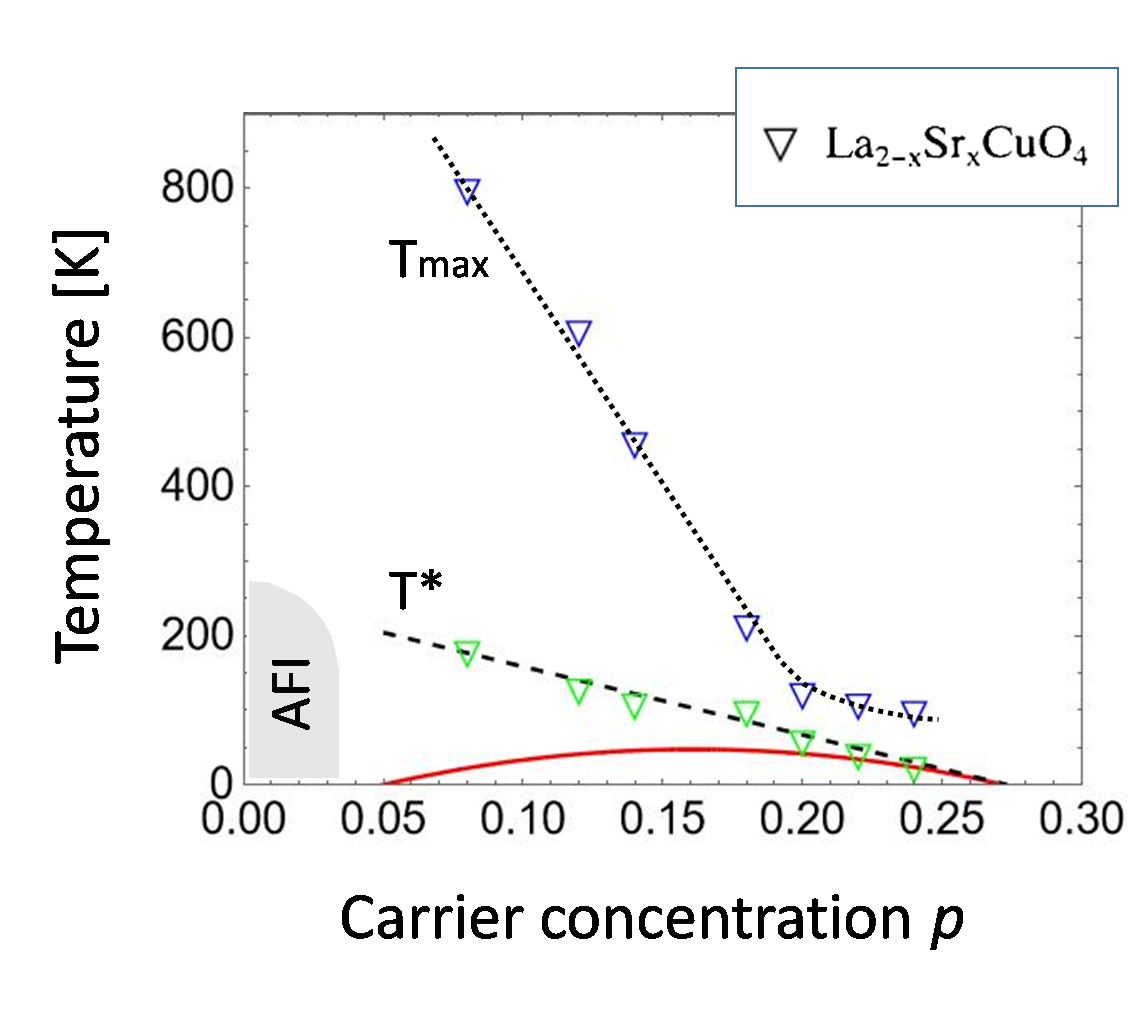}
\includegraphics[height=5.0 cm]{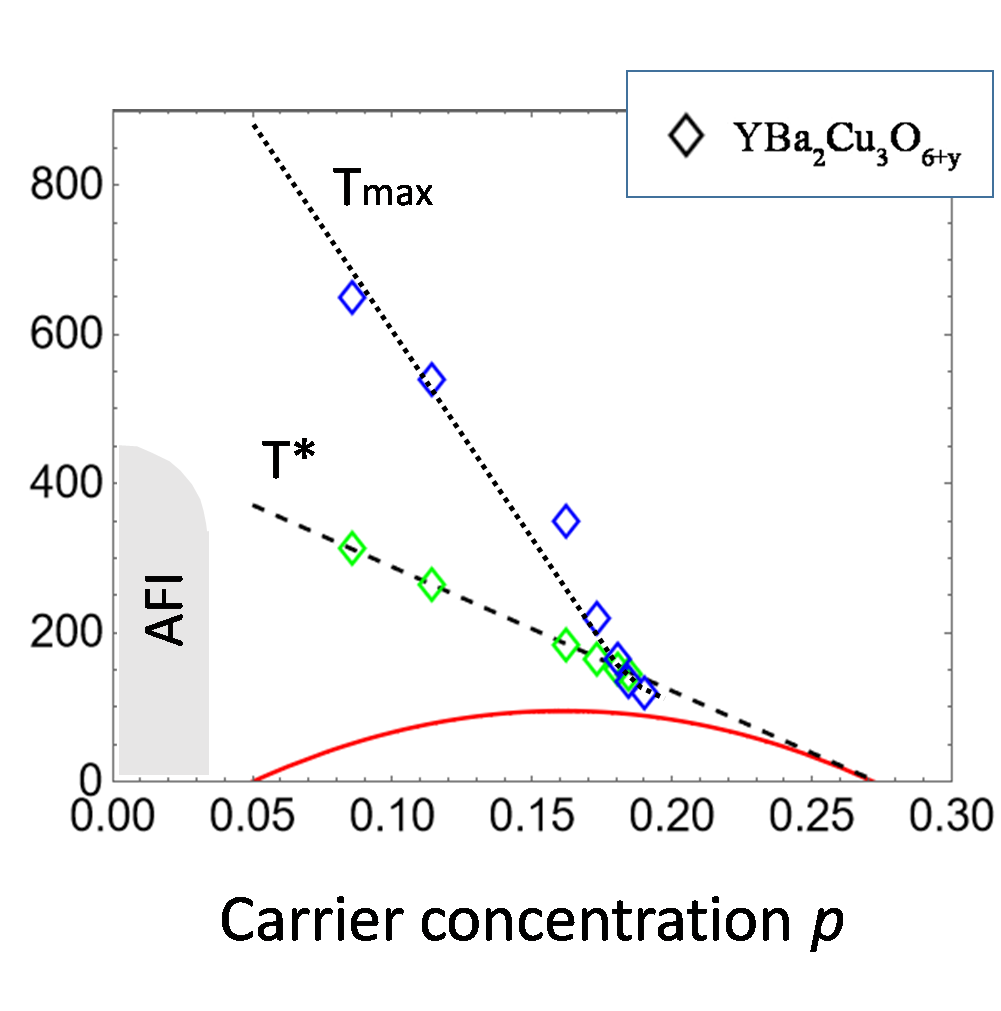}
\includegraphics[height=5.0 cm]{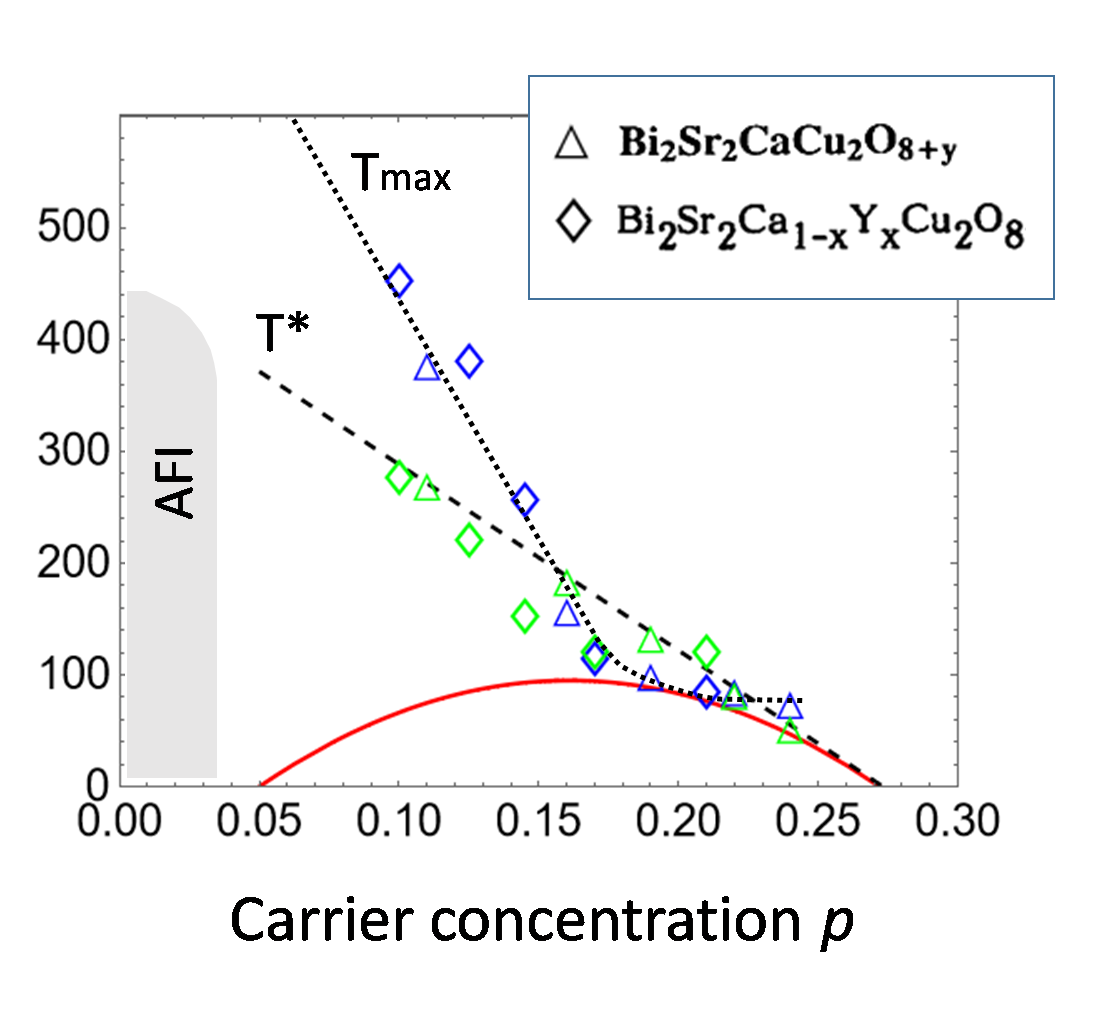}
\caption{(Color online) $T^*$ and $T_{max}$ as a function doping, as deduced from the fits for the four materials of Fig.\,\ref{Fig_Fits}. Note that both $T_{max}(p)$ and $T^*(p)$, for a wide range of $p$, display very similar laws. $T^*(p)$ runs tangential to the dome and seems to approach, at low doping, the N\'eel temperature of the AF insulator state (shaded area), as also noted in \cite{PRB_Cyr-Choiniere2018}. $T_{max}(p)$ remarkably follows a universal law for the four materials.} \label{Fig_Param}
\end{figure*}

The fit procedure contains the four terms of equation \ref{Equa_Chi_terms}, but the overall shape is clearly dominated by the magnetic plus the Pauli terms. The constant $\chi_0$ is fixed for each material for the lowest doping value (within a variation of 10 percent for Y-BSCCO).

We extract from the fits the values of the parameters, $T_{max}$, $T^*$ and the Pauli amplitude, as a function of doping.
Results are plotted in Fig.\,\ref{Fig_Param} for LSCO (left panel), YBCO (middle panel), for oxygen and Y-doped BSCCO (right panel). A striking feature is that, for all materials, $T_{max}(p)$ decreases linearly as a function of doping up to slightly overdoped regime ($p\simeq$0.19). This straight $T_{max}$ line extrapolates at $T=0$ to a doping value $p\simeq$0.23 for LSCO and YBCO and $p\simeq$0.20 for BSCCO. Towards the top of the dome, $T_{max}(p)$ has a more complex behavior deviating from linearity and saturating in the overdoped regime. These findings extend the work of Nakano et al. on LSCO \cite{PRB_Nakano1994}.

The pseudogap temperature $T^*$ deduced from the fits (see Fig.\,\ref{Fig_Param}, green dashed line) follows a straight line for LSCO, YBCO and BSCCO, as a function of doping, throughout the SC dome. For Y-doped BSCCO, small deviations from linearity arises near the top of the dome. In addition, for all the four compounds studied,  $T^*$  is higher than $T_c$ for all doping values and thus never crosses the $T_c$-dome. In conclusion, the $T^*(p)$ line extracted from the susceptibility curves closely matches the values obtained by ARPES \cite{PNAS_Vishik2012}. 
\begin{figure}
\includegraphics[width=8.4 cm]{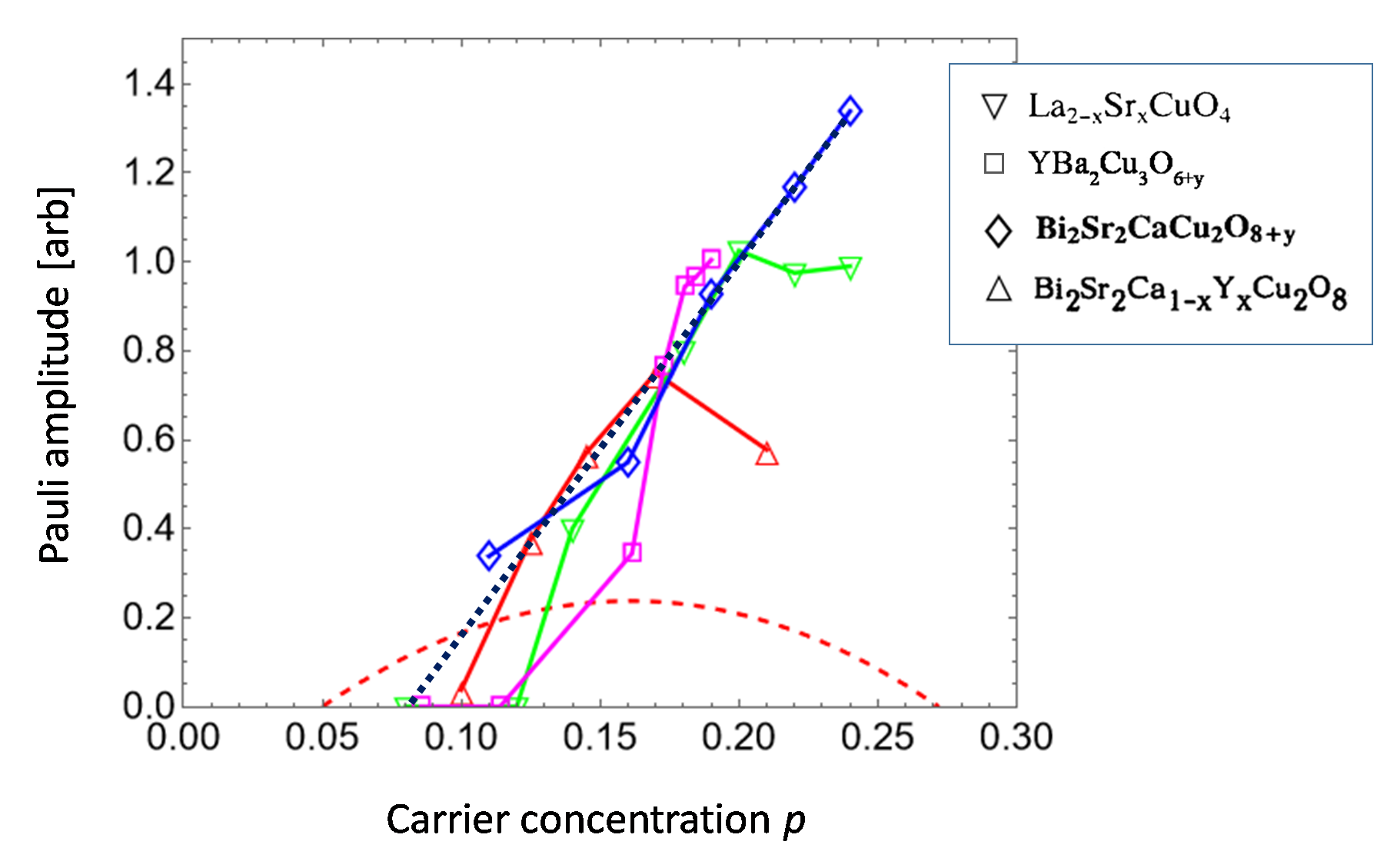}
\caption{(Color online) Pauli (pairon) amplitude as deduced from the
fits as a function of the carrier concentration. (Dotted line: $T_c$
dome for convenience). {The Pauli contribution to the susceptibility
is obtained by calculating the temperature-dependent DOS at the
Fermi energy within an energy window of the order $k_BT$ (see
Eq.\,\ref{Equa_Chi_Pauli}).} } \label{Fig_Amp_Pairons}
\end{figure}

The Pauli amplitude deduced from the fits is plotted in Fig.\,\ref{Fig_Amp_Pairons}. It is increasing as a function of $p$ for BSCCO and LSCO and YBCO. For Y-doped BSCCO, it is first monotonically increasing, but then decreases abruptly in the highly overdoped regime.
In spite of the relative uncertainty, our results for the amplitude are similar to the behavior of $\gamma_N(p)$, the gamma coefficient of the specific heat in the normal state \cite{PhysicaC_Loram1989,PhysicaC_Momono1994}. Indeed, for a standard metal, both $\gamma_N$ and $\chi_{Pauli}$ are proportional to the DOS at the Fermi energy.

\vskip 2mm

\subsection{Discussion}

\begin{figure}
\includegraphics[width=8.4 cm]{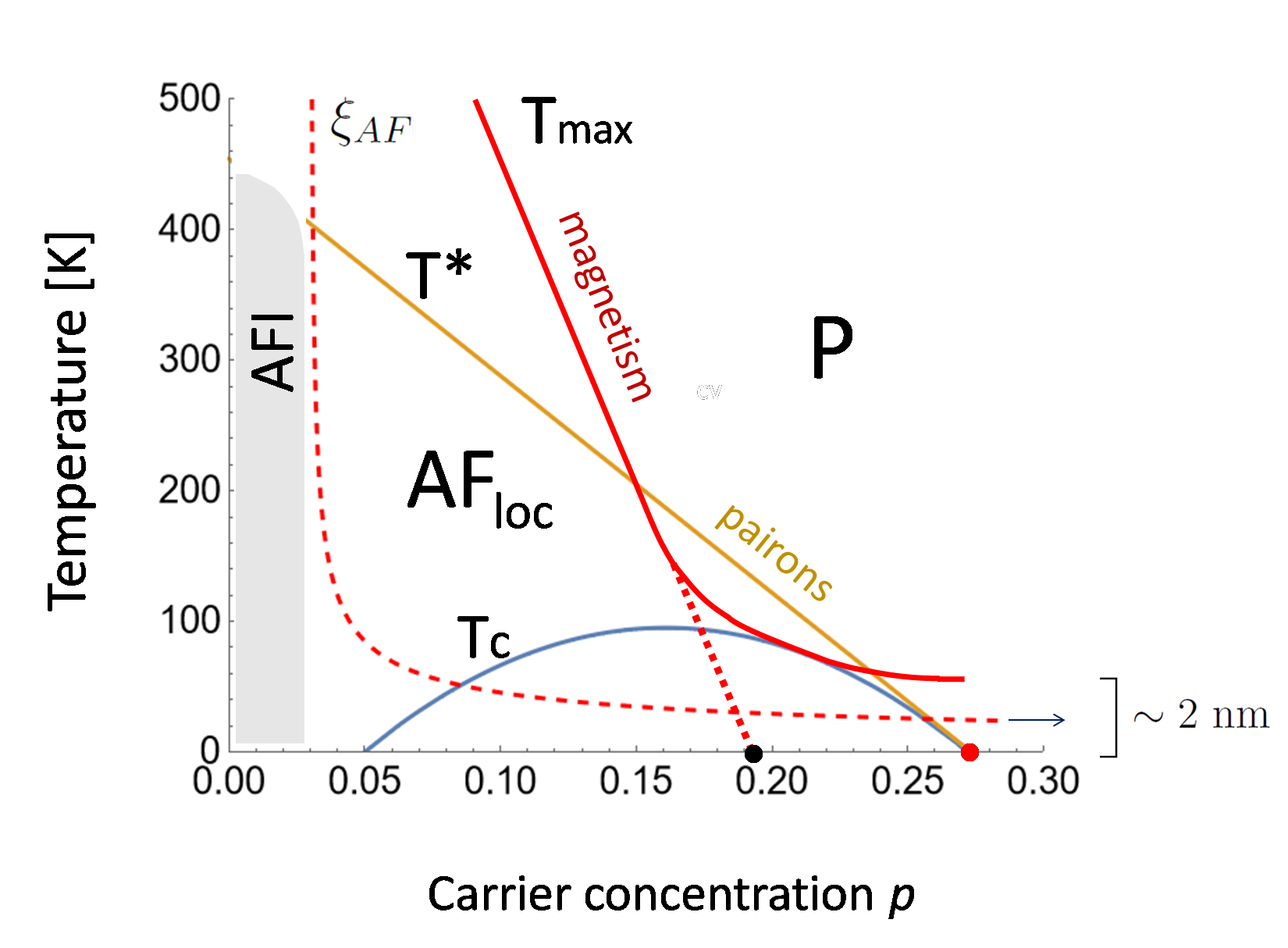}
\caption{(Color online) Phase diagram for BSCCO deduced from the
susceptibility data: $T_{max}$ (red line), $T^*$ (yellow line) as a
function of $p$, the hole concentration. The critical temperature
for BSCCO (blue line) and the antiferromagnetic correlation length
(dashed line) are also indicated. {$\xi_{AF}$ is approximated by the
average distance between holes, as found by Birgeneau et al.
\cite{PRB_Birgeneau1988}.} The $T_{max}(p)$ line separates the phase
diagram into two regions: on the left ($T<T_{max}(p)$) characterized
by local AF order (AF$_{loc}$) and on the right ($T>T_{max}(p)$),
characterized by strong magnetic fluctuations (P).}
\label{Fig_PhaseDiag}
\end{figure}
Two temperature scales emerge from our analysis,  $T^*$ the
temperature at which the pseudogap in the electronic DOS at the
Fermi level opens, and $T_{max}$ the characteristic temperature of
antiferromagnetic correlations. The $T^*(p)$ line is found to be
tangential to the superconducting dome, and therefore never crosses
the $T_c$ line. The magnetic scale $T_{max}$ behaves in a different
way: in the four materials we have considered, it decreases linearly
as a function of $p$, with a steeper slope, up to the slightly
overdoped value $p\sim$0.19. It then saturates in the overdoped
region, in agreement with the persistence of magnetic correlations
seen up to $p_c$ reported by
\cite{PRB_Wakimoto2005,PRL_Wakimoto2004,PRL_Wakimoto2007} and even
beyond \cite{Nature_Dean2013}. In addition, the extrapolation of the
linear behavior gives a critical value close to $p=$0.2 in the four
materials.

Although very prominent, this magnetic temperature scale has been improperly attributed to the pseudogap temperature. For example, the `pseudogap' temperature inferred from Hall measurement for LSCO is very close to our  $T_{max}(p)$. The confusion between  $T^*$ and  $T_{max}$  explains the contradiction between the phase diagrams deduced from susceptibility, transport measurements, on the one hand, and those deduced from spectroscopic measurements on the other hand.

Transport and susceptibility measurements are not direct probes of the DOS at the Fermi
energy since they are not only sensitive to mobile carriers at the Fermi level,
but also to their multiple diffusion processes. 
On the contrary, tunneling spectroscopy and ARPES directly probe the
quasiparticle {peaks} with a high precision. This might explain the
apparent contradiction between the two different pseudogap lines
found in the literature.

The interpretation of this new phase diagram within the pairon model
is at this point speculative, because of the absence of exact
solutions for microscopic models. Indeed, it is important to note
that another model with a similar temperature-dependent DOS  at the
Fermi energy, i.e. decreasing below a characteristic temperature
$T^*$, would have given qualitatively the same behavior for
$T^*(p)$. Therefore, one cannot exclude other interpretations for
the origin of the pseudogap. A pseudogap line tangential SC dome is
also found by Marino et al.
\cite{SUST_Marino2020,SUST_Res_Marino2021,SUST_ResMag_Marino2021},
with a model where the pseudogap is attributed to the condensation
of excitons.

In this work, we interpret the pseudogap as being due to the
formation of incoherent hole pairs or pairons. In this scenario,
there is a natural link between $T_{max}$ and $T^*$. Although having
a very different doping behavior, it is important to stress that
both temperature scales, $T_{max}$ and $T^*$, are proportional to
the same energy scale, the exchange energy $J$. For the undoped 2D
square lattice $T_{max}$ is given by \cite{JphysChemSol_Lines1970}:
\begin{equation}
k_BT_{max}\approx 1.12J\times S(S+1)
\end{equation}
whereas for $T^*$, a mean-field equation for pairons allows to write \cite{EPL_Sacks2017}:
\begin{equation}
2.2k_BT^*\approx J\left(1-\frac{p}{p_c}\right)
\label{Equa_Tstar}
\end{equation}
where $p_c$=0.27 is the doping for which superconductivity vanishes at the dome extremity (see Fig.\,\ref{Fig_Phase diagrams}, left panel). Note that for LSCO, the $T^*$ and $T_c$ are lower by a factor of two. The factor 2.2 in Eq. \ref{Equa_Tstar} should then be replaced by 1.1.
The validity of Eq. \ref{Equa_Tstar} for $T^*(p)$, as determined by electron spectroscopies, is illustrated Fig.\,\ref{Fig_AN_gap}. It is quantitatively compatible with the PG temperature deduced from the susceptibility data.

In addition, for the magnetic scale, we find the surprisingly simple relation
\begin{equation}
T_{max}(p)\approx 960\,K\left(1-5p\right)
\end{equation}
which extrapolates to zero at $p=$0.2.
These two simple linear laws for $T_{max}$ and $T^*$ give a slope difference of about a factor of two, except for LSCO, and a crossing point near the top of the dome. The above linear law for $T_{max}(p)$ is close to the energy gap $E_g/k_B$ deduced from both specific heat \cite{JPhysChemSol_Loram1998} and susceptibility \cite{PhysicaC_Naqib2007,ScSciTech_Naqib2008}.

A summary of our findings is presented in the phase diagram in Fig.\,\ref{Fig_PhaseDiag}.
The $T_{max}(p)$ line clearly separates the cuprate phase diagram in two main regions, to the left and to the right of the crossing point of $T_{max}(p)$ and $T^*(p)$. For $p\lesssim$0.16, the pseudogap temperature $T^*$ is smaller, or even much smaller, than $T_{max}$. This means that pairon formation occurs in the region of local AF order (indicated by AF$_{loc}$ in Fig.\ref{Fig_PhaseDiag}). Therefore, above $T_c$ pairons dissociate in a system with strong magnetic correlations.

On the other hand, to the right of the crossing point for $p>0.16$, $T_{max}$ and $T^*$ are close and pairons dissociate in a different magnetic environment (indicated by `P' in Fig.\ref{Fig_PhaseDiag}). Since we know from the fits that the susceptibility is close to a Curie law, we identify this region as `paramagnetic'. In this region, the spin-spin correlations are small ($T_{max}\sim 50-60$K) and the correlation length $\xi_{AF}$ is of the order of a few lattice constants \cite{PRB_Birgeneau1988}.

We see that crossing the top of the dome, from  left to right, there is a clear change of regime which should affect all physical quantities. First, the magnetic response changes from strong antiferromagnetic correlation (AF local) to weak correlations (P). Second, pairon decay into quasiparticles becomes much more dominant, leading to the increasing Fermi-arcs, as reported in the literature \cite{Nat_Norman_1998} and confirmed by the Pauli amplitude in Fig.\,\ref{Fig_Amp_Pairons}. Consequently, quasiparticles above $T_c$ must have a significantly different self-energy and spectral function, as possibly seen in \cite{Science_Chen2019}. This should affect the resistivity, specific heat and Knight shift.

Finally, as indicated in Fig.\,\ref{Fig_PhaseDiag}, in the overdoped regime both $T^*$ and $T_c$ approach each other and finally vanish at $p_c=0.27$, the critical point at the end of the superconducting dome. On the other hand, magnetic correlations persist in the overdoped regime \cite{Nature_Dean2013,PRB_LeTacon2013,PRB_Peng2018} and the magnetic scale $T_{max}$ remains finite.

\subsection{Conclusion}

In this paper, we have tackled the issue of two contradictory phase diagrams emerging from transport, specific heat, susceptibility on the one hand versus electron spectroscopies on the other hand.

To this end, we analyzed the magnetic susceptibility, $\chi(T)$, measured in high-$T_c$ cuprates in La$_{1-x}$Sr$_x$CuO$_4$, Bi$_2$Sr$_2$Ca$_{1-x}$Y$_x$Cu$_2$O$_8$, Bi$_2$Sr$_2$CaCu$_2$O$_{8+y}$ as well as the Knight shift measured in YBa$_2$Cu$_3$O$_{6+y}$.
Our analysis of $\chi(T)$ contains two major contributions, the magnetic response of the 2D AF lattice, and the Pauli term containing the Fermi level density of states. It is remarkable that the same function matches the measured $\chi(T)$ for all four materials from underdoped to overdoped regimes.
\vskip 2mm
To summarize the results:
\begin{enumerate}[label=\roman*)]
\item
Two temperature scales emerge from the analysis: $T_{max}$, the characteristic scale for antiferromagnetic correlations, and  $T^*$, characterizing the opening of the pseudogap in the electronic density of states at the Fermi level. The DOS was calculated in the framework of the pairon model, wherein hole pairs form due to their local antiferromagnetic environment below $T^*$. The measured Pauli term in $\chi(T)$ is therefore fully consistent with excited pairons existing above $T_c$.
\item
The magnetic contribution describes very well the overall shape of the susceptibility.
In particular, $T_{max}(p)$ decreases linearly over a wide range of hole doping $p$ but saturates in the overdoped regime. The extrapolated line ends at a critical doping $p\sim$0.2 for all the materials studied. Concomitantly, $T^*(p)$ is linear, with a smaller slope, and tangential to the superconducting dome, as confirmed by electron spectroscopies.
\item
In our concluding phase diagram, we identify two different regions, relative to the crossing point of $T_{max}$ and $T^*$. It implies a significant change of behavior in the magnetic correlations occurring from left to right at the top of the superconducting dome. Correspondingly, from underdoped to overdoped, we expect a qualitative change in the pairon dissociation, implying modifications in the quasiparticle spectral function.
\item
Our findings strongly suggest that the characteristic temperature
$T_{max}(p)$ inferred from susceptibility and transport measurements
has been incorrectly interpreted as the pseudogap temperature.
Furthermore, the linear behavior found for $T^*(p)$, indicates that
the same physics governs the {pseudogap formation} all along the
superconducting dome.
\item
We interpret the pseudogap as being due to the formation of
incoherent hole pairs, or pairons. Remarkably, {in this scenario},
the two temperature scales $T_{max}(p)$ and $T^*(p)$ coexist in the
same phase diagram and are proportional to the same energy, the
exchange energy $J$.
\end{enumerate}
\vskip 2 mm {\small Supplementary material, on the model and data
analysis, is available upon request from the corresponding
author$^{(1)}$}.

\subsection{Acknowledgments}

\end{document}